\documentclass[fleqn,10pt]{article}
\usepackage{graphicx}
\usepackage{authblk}
\usepackage{amsmath}
\usepackage{amssymb}

\usepackage[textwidth=16cm,textheight=24cm]{geometry}

\newcommand*{\TitleFont}{%
      \usefont{\encodingdefault}{\rmdefault}{b}{n}%
      \fontsize{13}{20}%
      \selectfont}

\title{\TitleFont{Experimental realization of the Yang-Baxter Equation via NMR interferometry}}

\author[1,2]{F. Anvari Vind}
\author[2]{A. Foerster}
\author[1]{I. S. Oliveira}
\author[1]{R. S. Sarthour}
\author[3]{D. O. Soares-Pinto}
\author[1]{A. M. Souza}
\author[1,*]{I. Roditi}

\affil[1]{\small Centro Brasileiro de Pesquisas F\'{\i}sicas, Rua Dr. Xavier Sigaud 150, 22290-180 Rio de Janeiro, RJ, Brazil}
\affil[2]{\small Instituto de F\'{\i}sica da UFRGS, Av. Bento Gon\c calves, 9500, Agronomia, Porto Alegre, RS, Brazil}
\affil[3]{\small Instituto de F\'{\i}sica de S\~{a}o Carlos, Universidade de S\~{a}o Paulo, CP 369, 13560-970 S\~{a}o Carlos, SP, Brazil}

\affil[*]{\small roditi@cbpf.br}
\date{}

\begin{document}

\flushbottom
\maketitle
\thispagestyle{empty}

\begin{abstract}
The Yang-Baxter equation is an important tool in theoretical physics, with many applications in different domains that span from condensed matter to string theory. Recently, the interest on the equation has increased due to its connection to quantum information processing. It has been shown that the Yang-Baxter equation is closely related to quantum entanglement and quantum computation. Therefore, owing to the broad relevance of this equation, besides theoretical studies, it also became significant to pursue its experimental implementation. Here, we show an experimental realization of the Yang-Baxter equation and verify its validity through a Nuclear Magnetic Resonance (NMR) interferometric setup. Our experiment was performed on a liquid state Iodotrifluoroethylene sample which contains molecules with three qubits. We use Controlled-transfer gates that allow us to build a pseudo-pure state from which we are able to apply a quantum information protocol that implements the Yang-Baxter equation.
\end{abstract}

%

\section*{Introduction}

In recent years, the Yang-Baxter equation (YBE)  \cite{Yang1,Gaudin,Baxter,qism1}, an important tool in theoretical physics, has attracted much attention in the context of quantum information science. It has been found that the YBE is closely related  to the generation of quantum entanglement \cite{Ge}. Furthermore a new quantum computation model based on the notion of integrability was proposed, where the quantum gates are related to unitary solutions of the YBE \cite{Nayak,KLom,Yzhang}. 

Formally, the YBE is expressed by, 
\begin{flalign}
\mathbf{R_{12}(v_1,v_2)R_{13}(v_{1},v_{3})R_{23}(v_{2},v_{3})=R_{23}(v_{2},v_{3})R_{13}(v_{1},v_{3})R_{12}(v_{1},v_{2})},
\end{flalign}

\noindent where $\mathbf{v_k}$, $k=1,2,3$, are called spectral parameters, or rapidities as they may also have a kinematical interpretation, and the $R$-matrix acts on a product space $\mathbf{V}\otimes\mathbf{V}$ \cite{perk}. The above equation provides a sufficient condition for quantum integrability and leads to  a consistent and systematic method to construct integrable models. 

In the quantum computing framework the R-matrix corresponds to a quantum gate \cite{KLom}, but this operator and the YBE may also have several other physical interpretations. In $(1+1)$-dimensional
quantum field theory/scattering theory the YBE means that the process of 
3-particle scattering is reduced to a sequence of pairwise collisions which do not depend on the 
time ordering of the 2-body collisions \cite{Zamolodchikov}. In this case $R$ is interpreted as the two-body scattering matrix (usually denoted $S$-matrix) and the Yang-Baxter Equation has the 
name of ``factorization equation". In vertex models of classical statistical physics, the
Yang-Baxter Equation appears as a condition for the vertex weights $R$ which allows for the 
exact solution of the corresponding model \cite{Baxter}. In this context it is usually referred as a 
``star-triangle" relation, interestingly inspired in a work of 1899 by a Brooklyn engineer, Kennelly, on electric networks,
using Kirchhoff's laws \cite{perk}. On a more recent note, there has been a considerable increase of investigation of these structures related to quantum integrability due to several new exact results that are playing an important role in the progress of our understanding of the AdS/CFT correspondence \cite{beisert}. It is worth mentioning, in addition, the interest raised by the realization of integrable systems, in ultracold physics. \cite{Kinoshita,Liaoetal}.

Another useful way to write the YBE is obtained by applying a permutation operator, $P$, to the $R$-matrix, such that,
$\check{R}=PR : \mathbf{V_a} \otimes \mathbf{V_b} \to \mathbf{V_b}\otimes\mathbf{V_a}.$
\noindent Assuming that the $R$-matrix satisfies the so called difference property \cite{Zamolodchikov, Wadati, Gomez} and by a simple transformation of the spectral parameters \cite{Gould1}, $x=e^u$ and $y=e^v$, it is possible to find a multiplicative form for the YBE,
\begin{flalign}\notag
\mathbf{\check{R}_{12}(x)\check{R}_{23}(xy)\check{R}_{12}(y)=\check{R}_{23}(y)\check{R}_{12}(xy)\check{R}_{23}(x)}.
\end{flalign}
The interest in this form comes from its relation to the braid group which has been recently linked to topological quantum computation \cite{Nayak,KLom,Yzhang} and a scheme for its verification through an optical setup has been proposed \cite{MolinGe} and achieved \cite{long}. 

Therefore, due to the importance of the YBE to physics and quantum information processing (QIP) and the increasing possibility of an experimental approach in solvable systems \cite{Guan}, direct quantum simulations of the YBE ought to be investigated. It is widely accepted in QIP that the development of large scale quantum processors depends broadly on two basic issues: i) the construction of systems containing a large number of qubits and (ii) the ability to control their quantum states, and implement correction protocols to prevent decoherence caused by unavoidable interactions with the environment. So far, NMR has been one of the leading techniques to demonstrate those aspects in small systems. In fact, the development of special pulse engineering techniques has allowed NMR to be applied to QIP with great success, with experimental demonstrations of different quantum protocols and algorithms in liquid and solid-state samples, including error-correction protocols \cite{Ivan1,Ivan2}. NMR quantum information processors provide a good testbed for QIP tasks, such as demonstrating quantum algorithms \cite{r1,r2,r3,r4}, fundamental physics studies such as delayed choice \cite{r5,r6}, quantum tunneling \cite{r7}, quantum dynamics \cite{r8} and PT symmetries \cite {r9}, and the present experiment in the Yang-Baxter equation. NMR implementations have, as well, already provided a certain number of observations linked in various ways to particular solvable systems\cite{Ivan2,Nano, Cappellaro, Zhangetal, feldman, araujo-ferreira, auccaise, laflamme, glaser}, and in a wider way to quantum integrable systems \cite{lloyd}. 

Here we follow the route of investigating this core relation behind quantum integrable systems, which we do by means of an NMR interferometric experiment as a tool to directly quantum simulate the YBE. The present demonstration of the YBE through QIP and NMR is the first one using this technique and opens up a way to implement quantum entanglement with integrability. \\

\section*{Results}

NMR implementations of quantum information processors are usually executed in an ensemble of identical and non-interacting molecules at room temperature, where nuclear spins are employed as qubits.  To implement quantum information in such systems, we need to prepare, from the thermal equilibrium state, the state:

\begin{equation}
\rho = \frac{1-\epsilon}{N} I + \epsilon \vert \psi \rangle \langle \psi \vert\,.
\label{pps}
\end{equation}
This state is a mixture of the pure state  $\vert \psi \rangle \langle \psi \vert$ and the maximally mixed state $I/N$, where $N$ is the dimension of the quantum system and $\epsilon \approx 10^{-5}$ is the thermal polarization of the system. Since the maximally mixed part does not produce observable signal, the overall NMR signal arises only from the pure state part $\vert \psi \rangle \langle \psi \vert$. Therefore the observed signal from a NMR system in the above state (\ref{pps}), called a pseudo-pure state (PPS), is equivalent to that from a system in a pure state, except that the PPS signal strength is reduced by a factor $\epsilon$.

We need to cast the YBE in a suitable form for our NMR verification. This can be done through its relation to braiding relations, emerging for instance from the exchange of anyons \cite{Nayak}, 
\begin{flalign}\label{braidgroup}
&\sigma_{j} \sigma_{k}\,=\,\sigma_{k} \sigma_{j}, |j-k|\ge2 \\\notag &\sigma_{j}\sigma_{j}^{-1}\,=\,\sigma_{j}^{-1}\sigma_{j}\\\notag
&\sigma_{j}\sigma_{j+1}\sigma_{j}\,=\,\sigma_{j+1}\sigma_{j}\sigma_{j+1}.
\end{flalign}
The operators $\sigma_{j}$ are the braid group generators. The YBE for $\check{R}$ matrices, in the so-
called braid limit, where $u=v$ and $|u|=\infty$, coincides with the braid relation in Eqs.
(\ref{braidgroup}). This is better seen using the notation $\mathcal{R}_j = \mathbb{I} \otimes ...\mathbb{I} \otimes \check{R} \otimes \mathbb{I}...\otimes \mathbb{I}$ acting on a product of vector 
spaces, $\check{R}$ acts on the spaces indexed $(j;j+1)$ and the identities on the other spaces, in the braid limit $\mathcal{R}_j $ furnishes a representation of the braid group \cite{Gomez}. The R-matrix that we need is obtained below.

First we remind \cite{MolinGe} that one can obtain two braid operations, $A$ and $B$, acting on a two-dimensional topological basis, such that $A$ acts as $\sigma_1$ and $B$ as $\sigma_2$, their respective matrix representation is
\begin{flalign}\label{AB}
&A=e^{-\dot{\imath}\frac{\pi}{8}}\left(\begin{array}{cc}1 & 0 \\0 & \dot{\imath}\end{array}\right);\:B=\frac{e^{-\dot{\imath}\frac{\pi}{8}}}{2}\left(\begin{array}{cc}1+\dot{\imath} & 1-\dot{\imath} \\1-\dot{\imath} & 1+\dot{\imath}\end{array}\right),
\end{flalign}
here, $i$ is the imaginary number. The braiding relation for these matrices is $ABA=BAB$, in the two-dimensional basis.

It is possible to generalize the above relation following a Yang-Baxterization \cite{Jones3,MolinGe2} procedure in order to introduce spectral parameters in a four-dimensional R-matrix that can be reduced to a two-dimensional one containing spectral parameters. It consists of writing the R-matrix as
\begin{flalign}
\check{R}(u)\,=\,a(u)\mathbb{I} + b(u) T,
\end{flalign}
more explicitly, in the usual notation,
\begin{flalign}
&\check{R}(u)_{12}\,=\,a_1(u)\mathbb{I} + b_1(u) T_{12}\\\notag
&\check{R}(u)_{23}\,=\,a_2(u)\mathbb{I} + b_2(u) T_{23},
\end{flalign}
\noindent and we are interested in the case $a(u)=a_1(u)=a_2(u),\,b(u)= b_1(u)= b_2(u)$ such that these scalar functions must be consistent with the YBE, also $T \equiv T_{jk}$ satisfies the so called Temperley-Lieb algebra \cite{TL} relations, $T^{2} \simeq T$, $T_{12}T_{23}T_{12}=T_{12}$ and $T_{23}T_{12}T_{23}=T_{23}$. The action of $\check{R}(u)$ on the two-dimensional basis \cite{MolinGe} leads to the definition of the spectral parameter dependent $A(u)$ and $B(u)$ by, respectively, the matrix elements of $\check{R}_{12}(u)$ and $\check{R}_{23}(u)$.

Consistency with the YBE provides $a(u)=\Gamma(u)$, a normalization factor, and $b(u)=\Gamma(u)[(2\sqrt{2}\dot{\imath}\zeta\beta u)/(1-2\dot{\imath}\zeta\beta u + \beta^{2}u^{2})]$ (where $\zeta=\pm 1$ and $\beta=-\dot{\imath}/c$, $c$ being the light speed). Then, introducing the transformation,
\begin{flalign}
\frac{1+2\dot{\imath}\zeta\beta u + \beta^{2}u^{2}}{1-2\dot{\imath}\zeta\beta u + \beta^{2}u^{2}}\equiv e^{-2\dot{\imath}\theta};\:\Gamma(u)\equiv e^{\dot{\imath}\theta},
\end{flalign}
one can write, in analogy with Eq. (\ref{AB}), two-dimensional matrices 
\begin{flalign} \label{AA}
&A(u)\equiv A(\theta) = e^{-i (2\theta) I_z}, \\\notag
&B(u)\equiv B(\theta) = e^{i (2\theta) I_x},
\end{flalign}
\noindent where $I_z=\frac{1}{2}\left(\begin{array}{cc}1 & 0 \\0 & -1\end{array}\right)$ and $I_x=\frac{1}{2}\left(\begin{array}{cc}0 & 1 \\1 & 0\end{array}\right)$, such that the operators are written in a form which is appropriate for NMR operations. Then the YBE we need is,
\begin{flalign} \label{yb2de}
A(\theta_1)B(\theta_2)A(\theta_3)\,=\,B(\theta_3)A(\theta_2)B(\theta_1).
\end{flalign}
The angle parameters which are the spectral parameters (or rapidities) must satisfy the following kinematical consistency relation
\begin{flalign} \label{angles}
\tan(\theta_2) \, = \, \frac{\sin(\theta_1 + \theta_3)}{\cos(\theta_1 - \theta_3)}.
\end{flalign}
Thus, in the NMR implementation, the operators (\ref{AA}) act on the ground states and excited states of the nuclear spin states, $|0\rangle$ and $|1\rangle$, respectively, which are used as qubits. In other words, one maps a two-dimensional invariant space on the two-level nuclear states of our sample.

Our experiment was performed on a liquid state Iodotrifluoroethylene sample. This contains molecules with three qubits, which are encoded in the $^{19}F$ spin-$1/2$ nuclei (see Figure \ref{iodotrifluoroethylene}). The phase decoherence times $(T_{2})$ for $F_1$, $F_2$ and $F_3$ are approximately 0.08s, 0.09s and 0.08s, respectively. The first step of the experiment consists of preparing, from thermal equilibrium, the PPS (\ref{pps}) with $\vert \psi \rangle = \vert 000 \rangle$, which correspond to a situation where all spins are on its ground state, for a PPS preparation we use the Controlled-transfer gates technique  \cite{jones}.

\makeatletter
\newenvironment{figurehere}{\def\@captype{figure}}{}
\begin{figurehere}
\begin{center}
\includegraphics[width=4.5in]{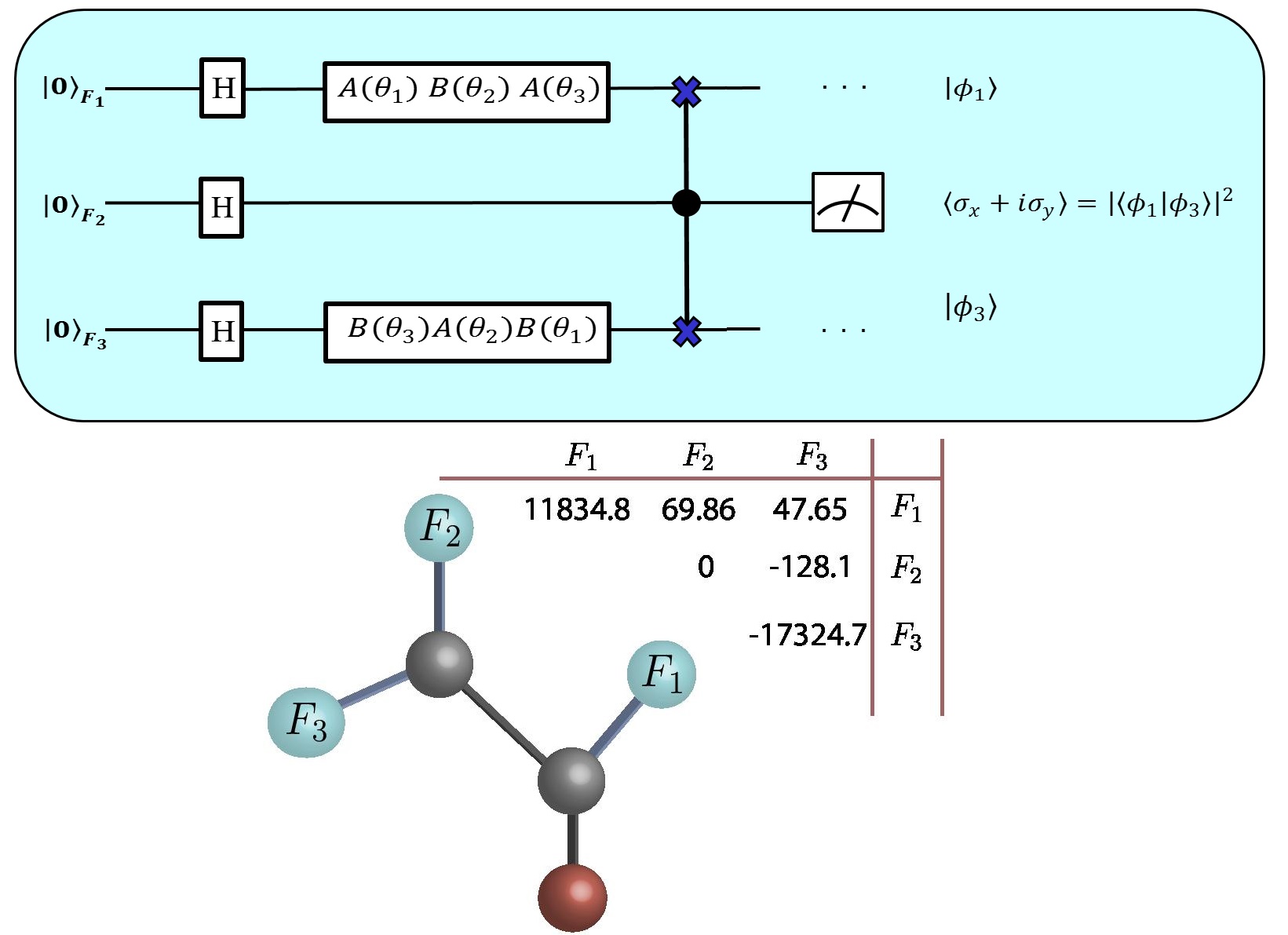}
 \end{center}
\caption{Quantum circuit diagram for implementation of the YBE as in \ref{yb2de} and \ref{angles}, where the operators $A$ and $B$ are the R-matrices, here the Yang-Baxterized braid operators acting on a two-dimensional basis. H is the Hadamard gate. The left hand side and right hand side of the YBE are applied on qubits 1 and 3, respectively, while qubit 2 is auxiliary. The YBE is satisfied when the overlap ${|\langle \phi_1 | \phi_3 \rangle|}^2$ equals 1. $F_1$, $F_2$ and $F_3$ are denoted as qubit 1, qubit 2 and qubit 3, respectively. The structure and parameters of the fluorine labeled Iodotrifluoroethylene molecule are also shown. The diagonal terms in the table are the chemical shifts (in Hz) of the fluorines. The off-diagonal terms are the coupling constants, also in Hz. The grey spheres represent carbon nuclei while the red one is the iodine.}
\label{iodotrifluoroethylene}
\end{figurehere}

After the initialization in the PPS we can perform the experiment for the verification of the Yang-Baxter Equation, the scheme of the experiment is shown in Figure \ref{iodotrifluoroethylene}. In the first step a $ \pi / 2$ pulse about $y$ axis is applied on qubits one and three, yielding to the three-qubit state $\vert \psi \rangle = \frac{1}{\sqrt{2}} (\vert 0 \rangle + \vert 1 \rangle) \otimes \vert 0 \rangle  \otimes \frac{1}{\sqrt{2}} (\vert 0 \rangle + \vert 1 \rangle)$.  In the next step, the Left-Hand-Side and the Right-Hand-Side of the 2D YBE (\ref{yb2de}) are applied on qubits one and three, respectively. By using Eq. (\ref{AA}), after some  manipulation the  2D YBE  can be brought to a sequence of single spin rotations, which is the sequence implemented in the experiment.

The qubits one and three in the output states are
$| \phi_1 \rangle$ and  $ | \phi_3 \rangle$, respectively (see Figure \ref{iodotrifluoroethylene}).  To verify the YBE we need to measure the overlap ${|\langle \phi_1 | \phi_3 \rangle|}^2$, if this quantity is equal to 1, then  $|\phi_1 \rangle = | \phi_3 \rangle$ and the YBE is satisfied, otherwise the YBE is not satisfied. To perform such verification we explore a quantum interferometric approach based on the Controlled-SWAP gate that can extract properties of quantum states without quantum tomography \cite{Arturyb}. In this approach (see the final step in Figure \ref{iodotrifluoroethylene}) the second qubit of our system is taken as a 
auxiliary qubit. After transforming the auxiliary qubit to the state  $\vert 0 \rangle + \vert 1 \rangle$, by a $ \pi / 2$ rotation about $y$ axis, a Controlled-Swap gate with the second 
qubit as control is applied. When a measurement is performed on the auxiliary, its normalized complex magnetization on the plane is directly related to the overlap between the states of the swapped qubits $\langle \sigma_{x}^{2} + i \sigma_{y}^{2} \rangle ={|\langle \phi_1 | \phi_3 \rangle}|^2\;$ \cite{Arturyb}.

\begin{figurehere}
\begin{center}
\includegraphics[width=3.5in]{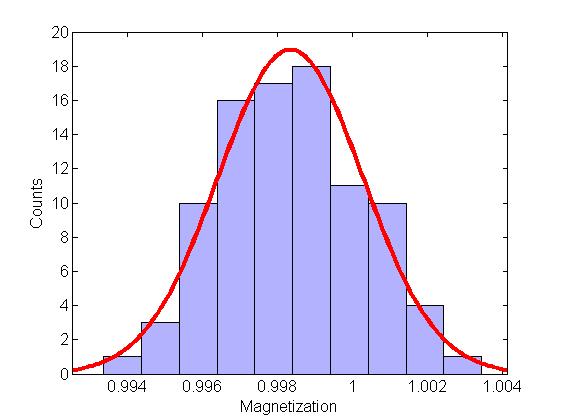}
 \end{center}
\caption{Total magnetization of qubit 2 for $\theta_1 = - \theta_3$, $\theta_2 = 0$ and $\theta_3$ varying from $0$ to $2 \pi$. The histogram displays the distribution of the normalized total magnetizations of the auxiliary qubits. The average magnetization is $0.998 \pm 0.001$, showing the expected result for the YBE.}
\label{correctangle}
\end{figurehere}

In Figure \ref{correctangle} we show the results for $\theta_1 = - \theta_3$, $\theta_2 = 0$ and $\theta_3$ varying from $0$ to $2 \pi$, in all cases the relation in (\ref{angles}) is satisfied. The histogram displays the distribution of the normalized total magnetizations of the auxiliary qubits. An average magnetization is $0.998 \pm 0.001$, showing the validity of the 2D YBE with good agreement.

\begin{figurehere}
\begin{center}
\includegraphics[width=5.5in]{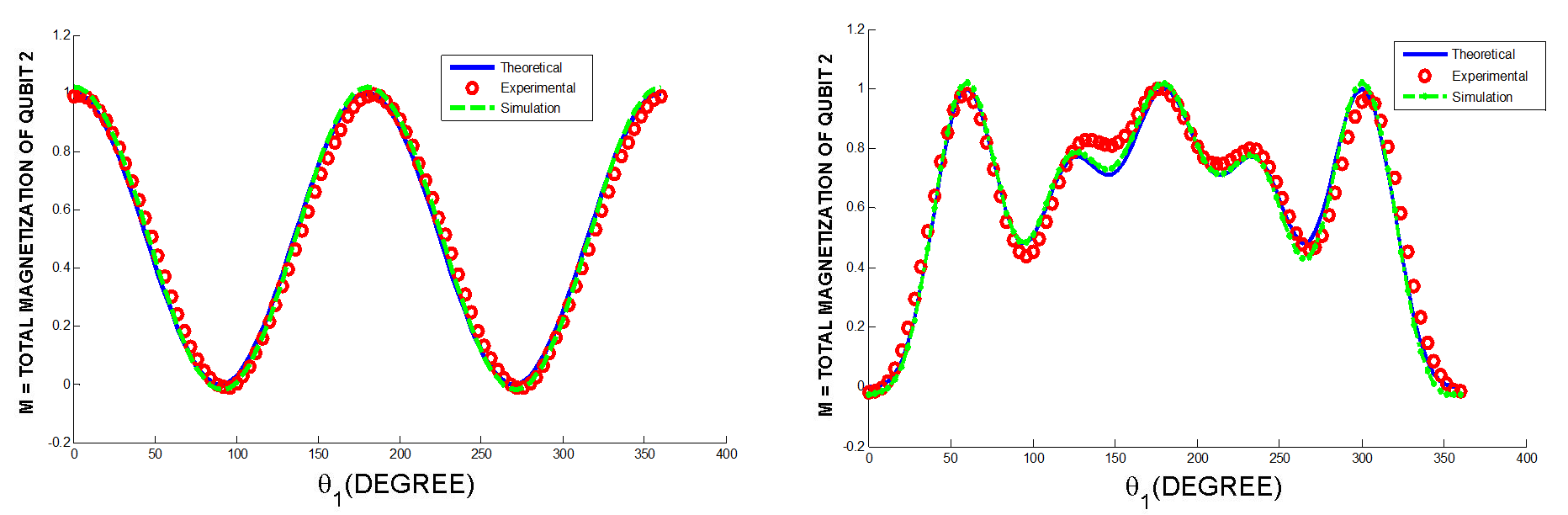}
 \end{center}
\caption{(a)  (On the left) Total magnetization of qubit 2 for $\theta_2 =  \theta_3 = 0$ and $\theta_1$ varying from $0$ to $2 \pi$. The blue line is the theoretical prediction, the green dashed line is a simulation of the experimental data and the red circles are the actual experimental results. For angles that do not satisfy the consistency relation (Eq. (\ref{angles})), the total magnetization of qubit 2 is under unity for the correct angles, the total magnetization of qubit 2 equals 1. (b)  (On the right) Total magnetization of qubit 2 for $\theta_1 = 2 \theta_2$, $ \theta_3 = \frac{\pi}{2}$ and $\theta_1$ varying from $0$ to $2 \pi$.}
\label{angleT}
\end{figurehere}

We also explore the cases where (\ref{angles})  is not satisfied. In Figure \ref{angleT}(a) we show the results for  $\theta_2 = \theta_3 = 0$ and $\theta_1$ varying from $0$ to $2 \pi$, these angle parameters do not satisfy the relation in (\ref{angles}) except for $\theta_1 = 0, \pi$ and $2 \pi$.  In Figure \ref{angleT}(b) we show the case where $\theta_1 = 2 \theta_2$, $ \theta_3 = \frac{\pi}{2}$ and when $\theta_1$ changes between $0$ and $2 \pi$. In this case YBE is not satisfied except for $\theta_1 = \frac{\pi}{3}, \pi, \frac{5 \pi}{3}$.

\section*{Discussion}

Using NMR techniques developed in the realm of quantum information processing we were able to experimentally verify the YBE. In order to achieve this we map a two-dimensional invariant space, related to a topological basis and anyon behavior, on the two-level nuclear states of our sample. In our verification we explored a quantum interferometric approach based on the Controlled-SWAP gate, a procedure that can extract properties of quantum states without quantum tomography. The present two-dimensional approach to investigate the YBE by means of NMR opens up a way to realize the four-dimensional case, which, differently from the present case, will necessarily involve Bell states, and reveal the possibility to implement quantum entanglement with integrability.
The issue of NMR entanglement is quite interesting and has been broadly discussed in the literature \cite{Ivan1}. Since the results of \cite{braunstein}, NMR has demonstrated in a series of experiments  its capability to produce quantum correlations, including entangled behaviour of qubits and their importance in quantum metrology \cite{girolami} and quantum simulations \cite{alexandre} have been investigated. The control over state creation and unitary manipulation, combined with the possibility of extending the number of qubits \cite{laflamme} unlocks new possibilities to the applications of NMR-QIP to the study of YBE. Therefore, in order to carry out the four-dimensional case in an NMR setup one needs to deal with a higher number of qubits. We are currently doing investigations in this direction and we believe that, as in the present work, the interweaving between quantum integrability and quantum information, as well as its relation with many-body, quantum fields and statistical physics, will unveil new and interesting patterns.

\section*{Methods}

For the Yang-Baxter experiment presented in this work we have used a  Varian $500$ MHz Spectrometer and a double resonance probe-head equipped with a magnetic field gradient coil. The three spin-$1/2$ $^{19}F$ nuclei of Iodotrifluoroethylene ($C_{2}F_{3}I$) molecule, dissolved in deuterated acetone, were used as qubits. The experiment has been performed at room temperature. The nuclear spins of the fluorine interact with the static magnetic field and with each other via a Ising-like model. The natural Hamiltonian of our system is described by
\begin{equation}
H = \sum_{j}\hbar\omega_{j}I_{z}^{j} + \sum_{j\neq k}\hbar 2 \pi J_{jk} I_{z}^{j}\otimes I_{z}^{k} + H_{RF}(t) ~,
\end{equation}
where $I_{z}^{j} = \sigma_{z}^{j}/2$ and $\omega_{j}$ being, respectively,  the nuclear spin operator in $z-$direction and  the Larmor frequency of the spin $j$ ($\sigma_{z}^{j}$ is the Pauli matrix); $J_{jk}$ are the couplings. $H_{RF}(t)$ is the radio-frequency (rf) Hamiltonian employed to control the qubits. The physical parameters of our molecule are shown in Figure\ref{iodotrifluoroethylene}.

Now we display in more detail the complexity of the sequence of unitary operations used in the experiment and shown in the quantum circuit in Figure \ref{iodotrifluoroethylene}, where time runs from left to right. In order to prepare the initial state and run the protocols necessary for the experiment we have used the NMR tools that were available. They were the  rf pulses and pulsed field gradients. The later is a gradient magnetic field applied along the $z$ direction causing an inhomogeneity in the static magnetic field. This gradient removes the coherences leading the system to a mixed state. From that new coherent states can be prepared by applying rf pulses. To prepare the initial state we have used the  Controlled-transfer gates methods \cite{jones}, the pulse sequence is presented in Figure \ref{initialstate}.

\begin{figurehere}
\begin{center}
\includegraphics[width=6.0in]{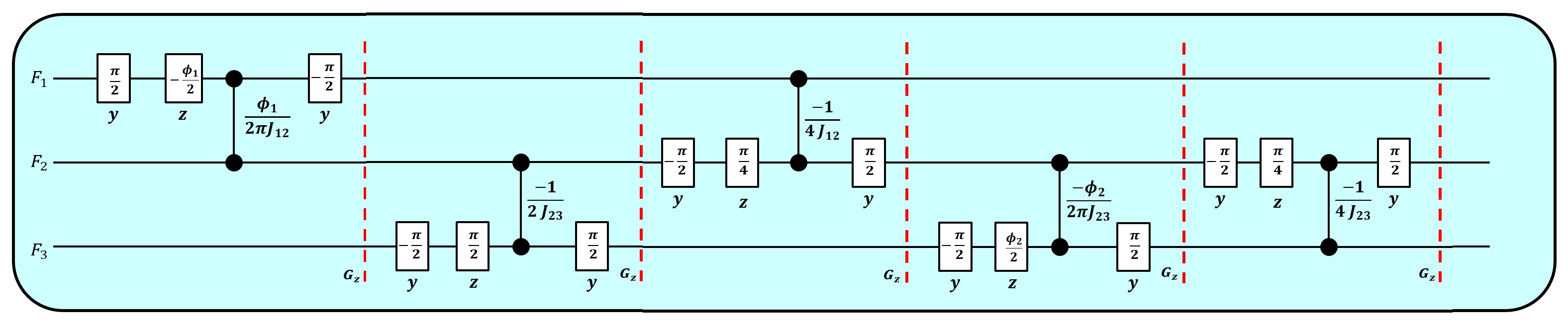}
 \end{center}
\caption{Pulse sequence for the initial state preparation. The initial state was prepared using Controlled-transfer gates methods. The boxes indicate the pulses that implement rotations applied to invidual qubits. The angles and phases of these rotations are indicated inside and below the boxes, respectively. Refocusing pulses are not shown. The free evolutions are represented by black dots connected by lines where the interaction of the two qubits, indicated by the dots, took place, for the time shown in the figure. The dotted red lines indicate when the field gradients were applied. $\phi_1 = 98.2\,^{\circ}$ and $\phi_2 = 135.59\,^{\circ}$.}
\label{initialstate}
\end{figurehere}

The Yang-Baxter protocol  was implemented using the sequence presented in Figure \ref{yangbaxterps}. For the Controlled-SWAP gate at the end of the quantum circuit we have used the following pulse sequence  presented in Figure \ref{conswap}.

\begin{figurehere}
\begin{center}
\includegraphics[width=3.5in]{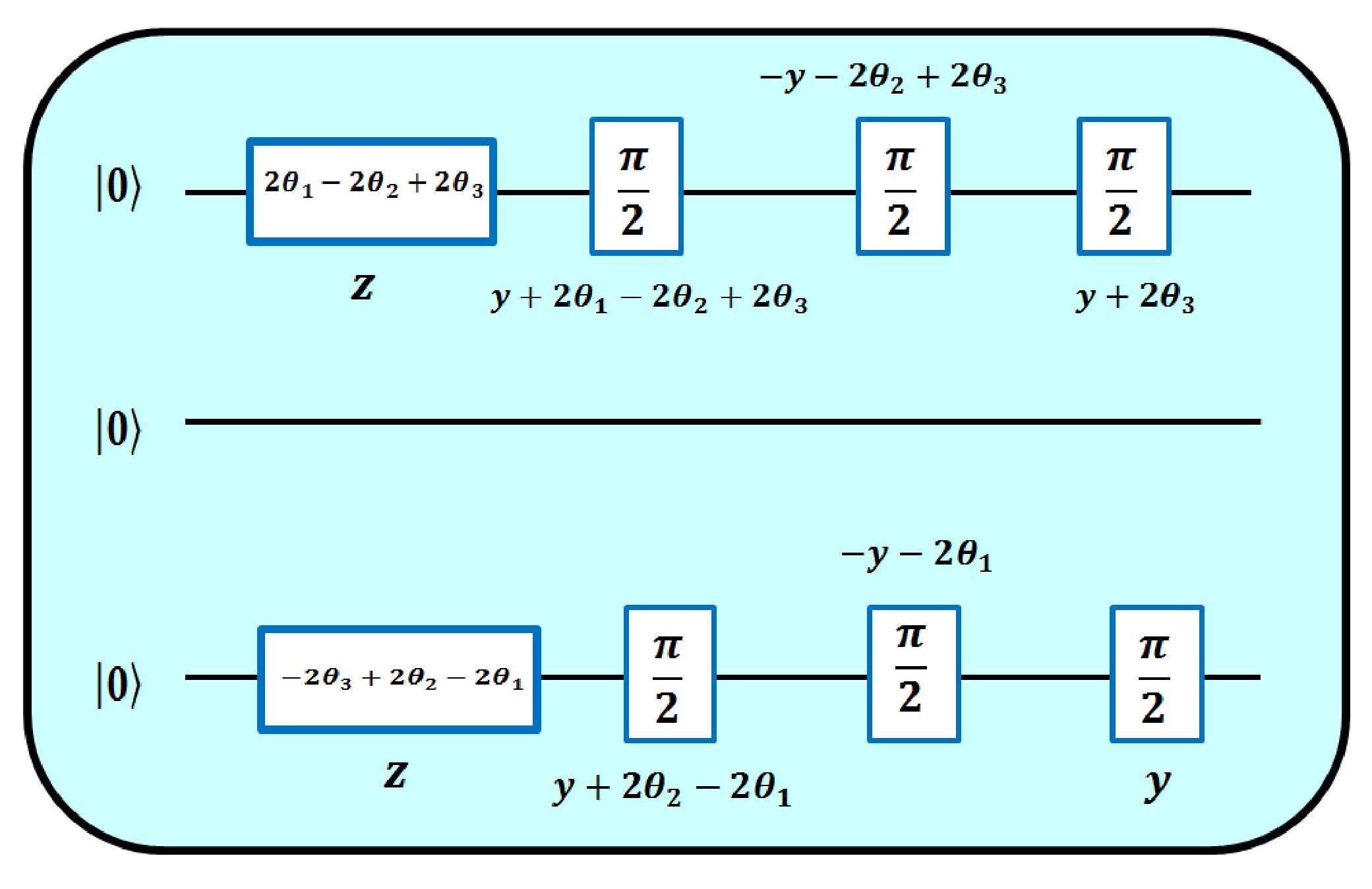}
 \end{center}
\caption{Yang-Baxter protocol. Sequence used to implement the YBE. The Left-Hand-Side and the Right-Hand-Side of the 2D YBE (\ref{yb2de}) are applied on qubits one and three, respectively. The boxes indicate the rotations applied to invidual qubits according to the kinematical consistency relations. The angles and phases of these rotations are indicated inside and below the boxes, respectively. Afterwards one has to use the Controlled-SWAP to check the overlap ${|\langle \phi_1 | \phi_3 \rangle|}^2$.}
\label{yangbaxterps}
\end{figurehere}

\begin{figurehere}
\begin{center}
\includegraphics[width=6.0in]{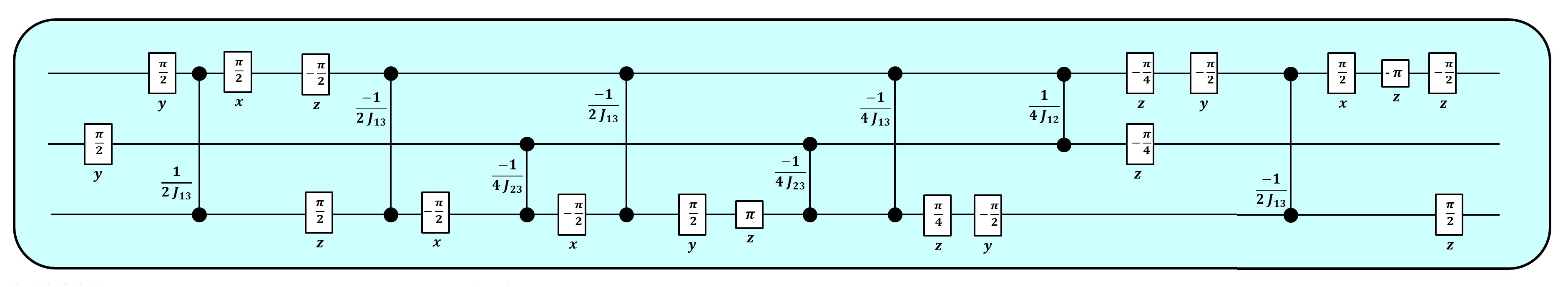}
 \end{center}
\caption{Controlled-SWAP. Quantum interferometric approach based on the Controlled-SWAP gate that provides the measure of the overlap ${|\langle \phi_1 | \phi_3 \rangle|}^2$. The boxes represent rotations applied to invidual qubits. The angles and phases of these rotations are indicated inside and below the boxes, respectively. Refocusing pulses are not shown. The free evolutions are represented by black dots connected by lines where the interaction of the two qubits, indicated by the dots, took place, for the time indicated in the figure.}
\label{conswap}
\end{figurehere}

To implement the operations we exploit standard Isech shaped pulses and numerically optimized GRAPE pulses \cite{Khaneja}. The GRAPE pulses are optimized to be robust to Radio-Frequency inhomogeneities and chemical shift variations.  For combining all operations into a single pulse sequence we have used the techniques described in \cite{Ryan,Bowdrey}. We built a computer program, similar to the NMR quantum compiler used in the 7 qubits NMR experiments \cite{Knill,Souza,Zhang}. The program minimizes the effects of finite pulsewidth, off-resonance errors  and unwanted coupling evolutions.


\section*{Acknowledgements}

The authors thank X.-W. Guan for many crucial suggestions.
The following agencies -- CNPq, CAPES, FAPERJ and INCT-IQ -- are gratefully acknowledged for financial support
\section*{Author contributions statement}

All authors contributed to the manuscript final text. F.A.V. and A.M.S. optimized  the pulses sequences, conducted the experiment and prepared the figures; I.S.O, R.S.S. and D.O.S.P.  participated on the experimental planning and discussion of results; A.F. and I.R. developed the theoretical ideas and discussed the results of the experiment, I.R. coordinated the project.
\section*{Additional information}

\textbf{Competing financial interests}

The authors declare no competing financial interests. 

\end{document}